\documentclass[aps,prl,twocolumn,showpacs,floatfix]{revtex4-1}

\usepackage{color}
\usepackage{bibentry}
\usepackage{amsmath,amssymb}
\usepackage{bm}
\usepackage{verbatim}
\usepackage{graphicx}
\usepackage{dcolumn}
\usepackage{amstext}
\usepackage{mathptmx}
\usepackage{verbatim}
\usepackage[colorlinks,urlcolor=black,citecolor=blue,bookmarks]{hyperref}

\usepackage{xcolor}
\usepackage{ulem}
\usepackage{color,soul}

\usepackage{subfigure}
\DeclareMathAlphabet{\mathpzc}{OT1}{pzc}{m}{it}

\begin{document}

\title{Hall anomaly and vortex charge in Bi$_2$Sr$_2$CaCu$_2$O$_x$}

\author{Yuval Nitzav}
\affiliation{Physics Department, Technion-Israel Institute of Technology, Haifa 32000, Israel.}
\author{Amit Kanigel }\thanks{Corresponding author: kanigel.amit@gmail.com}
\affiliation{Physics Department, Technion-Israel Institute of Technology, Haifa 32000, Israel.}

\date{\today}
\begin{abstract}
We present a systematic study of the Hall conductance in Bi$_2$Sr$_2$CaCu$_2$O$_x$ (Bi2212) thin films over a large range of doping. We find that in a large part of the phase diagram the Hall coefficient changes sign as a function of temperature in the flux-flow regime. By comparing data from many samples, we show that the sign reversal is tied to the superconducting transition and is not a result of a competing order. We then compare our data to the predictions of the Bardeen-Stephan model and show that in all  samples there is an additional negative contribution to the Hall conductivity. We extract from the negative excess Hall a vortex-charge that is found to be strongly doping dependent.
\end{abstract}

\maketitle

\section{Introduction}
The phase diagram of the cuprates is characterized by several competing and intertwined orders. In most of the doping range, 
the superconductivity dome obscures other phases at low temperature. By applying very high magnetic fields one can suppress superconductivity and reveal the underlying "normal" state.
Hall measurements in magnetic fields of up to 88T in YBa$_2$Cu$_3$O$_y$ (YBCO) have found that at low doping the Hall number is positive, at intermediate doping ($0.08 < p <0.16$) the Hall  number is negative and for $p>0.16$ the Hall number is positive again \cite{Badoux16,Leboeuf2011,LeBoeuf2007,Ando2004}. The negative Hall sign in YBCO at low temperatures \cite{LeBoeuf2007} is consistent with a Fermi surface reconstruction\cite{Harrison2012,Review_Sebastian_Proust,Leboeuf2011} caused by a charge density wave (CDW) \cite{NMR_CDW, RIXS_CDW,XRD_CDW}. 
The correlation length of the CDW increases as the superconductivity is suppressed by increasing the magnetic field \cite{CDW_Correlation_length}, suggesting that the CDW is "field-revealed" rather than "field-induced".  CDW instabilities were found  in  Bi$_2$Sr$_2$CaCu$_2$O$_y$ (Bi2212) \cite{dasilvaneto14} as well.

\begin{figure}[h]
\includegraphics[width=0.8\linewidth]{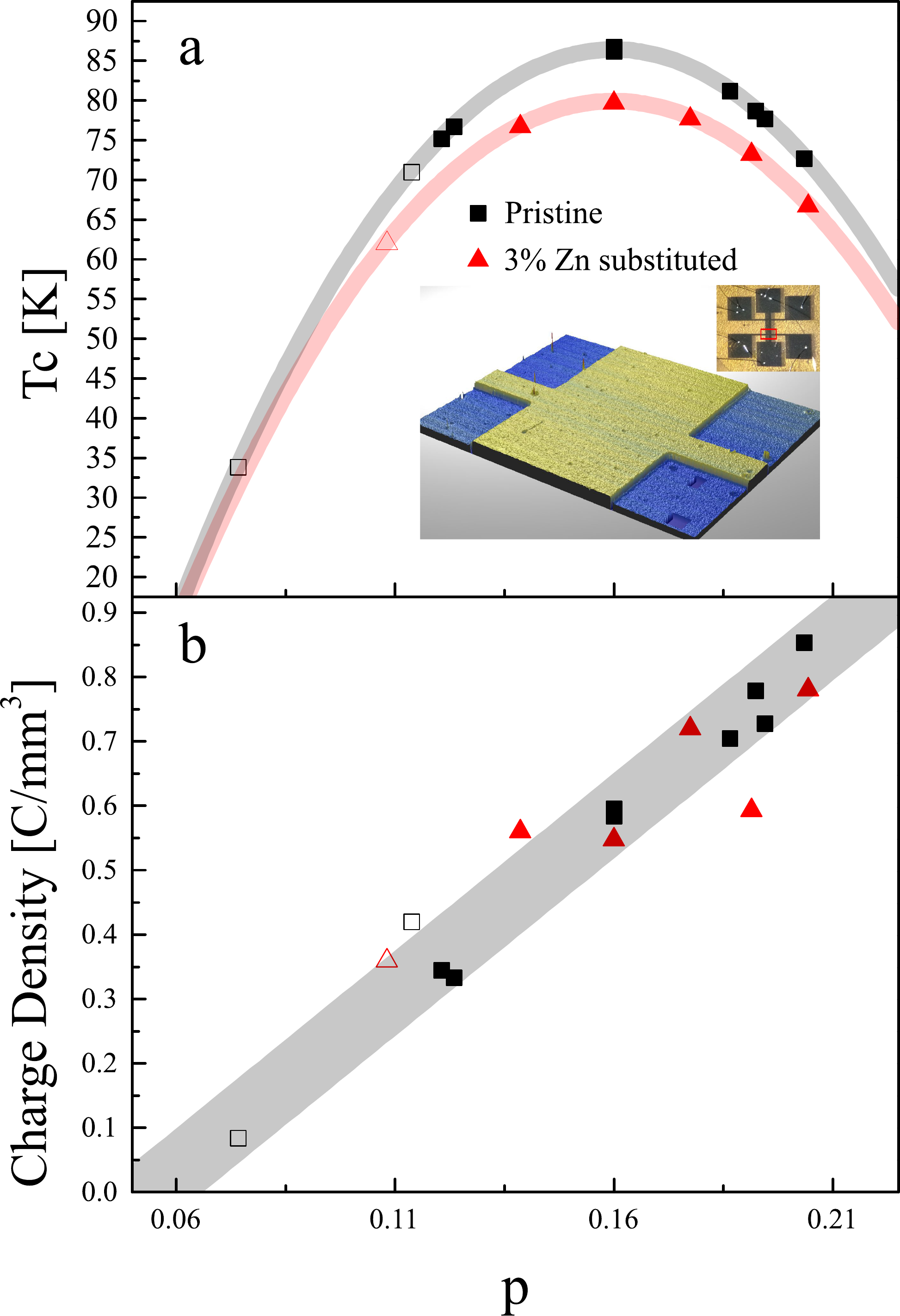}
\caption{(a) Critical temperature vs doping. In red are 3\% Zn doped samples, in black are pristine Bi2212 samples. Empty markers identify samples with no Hall-anomaly. In the inset we show an image of the Hall bar and a plot of the topography of part of the Hall bar as measured using an optical profilometer.
(b) Charge carrier density inferred  from the Hall resistance at 120K, as a function of doping - derived from the critical temperature and maximal critical temperature using the Presland formula \cite{PRESLAND199195}}
\label{fig1}
\end{figure}

There is an alternative way to learn about the state hiding beneath the superconducting dome. This can be done by measuring the resistivity in the flux-flow regime. It was shown by Bardeen and Stephen (BS) \cite{bardeen1965theory} and by Nozieres and Vinen \cite{Nozieres1966} that the resistivity in the flux flow regime is proportional to the vortex core resistivity. Since superconductivity is suppressed in the vortex cores, it is possible that the same Fermi surface reconstruction will also be present in the vortex cores.  This should result in a sign change of the Hall resistance in the flux flow regime.

Such a sign reversal, dubbed a Hall anomaly, has been the subject of many studies since the discovery of the cuprates. Hall measurements in YBCO \cite{Hagen1990,Galffy1988}, Bi2212 \cite{Iye1989}, and ErBa2Cu3O7\cite{Iye1989}, have shown a sign reversal near the critical temperature, indicating that the Hall anomaly is a universal property of the cuprates\cite{Nagaoka1998a}.
The sign change have been attributed to charged vortices \cite{khomskii1995charged,Kato1999b,Auerbach2019,FEIGELMAN1994VortexCharge}, additional forces acting on the vortices \cite{Kopnin1996a,Kolacek2000}, thermal fluctuations\cite{Wang1994a}, back-flow induced by hydrodynamic effects\cite{Ferrell1992b}, and a Magnus force which could be demonstrated to exist using a Berry phase argument\cite{ao1993berry}.
Recently, a double sign-reversal has been measured also in atomically thin Bi2212 samples. The sign reversal has been observed below, and slightly above the critical temperature\cite{Zhao2019}.
Whether the Hall anomaly is a result of a Fermi surface reconstruction due to a competing order, and its connection to the behaviour of the Hall number at high fields are still open questions. 

   In this paper, we report measurements of the conductivity (both longitudinal and Hall) of a series of Bi2212 thin films in a large doping range. We find a reversal of the sign of the  Hall number around the critical temperature for part of the samples.  
We examine the doping dependence of the Hall anomaly in Bi2212 over a large doping and critical temperature (T$_c$) ranges.  We use Zn substitution to decouple the critical temperature from the doping level. The substitution of a small amount of Cu atoms by Zn atoms lowers the critical temperature while effectively keeping the doping level unchanged \cite{Xiao1988a,Lubashevsky2011}.   This allows us to test for correlations between the Hall anomaly and the critical temperature and doping level independently.  
In addition, in YBCO it was shown that Zn substitution  reduces the CDW correlations substantially \cite{Keimer_zn_CDW}, so the Zn samples are expected to have a weaker Fermi-surface reconstruction.

\section{Experimental details}
A series of 200 nm thick films of pristine and 3\% Zn-substituted Bi2212 samples were prepared using a DC sputtering system on LaAlO$_3$ substrates. The doping level was set by annealing the samples at different temperatures and oxygen pressures.
Using deep UV lithography the films were etched into 6-contact Hall bars 1.5mm long and 140$\mu$m wide.  
 
In this configuration, we were able to measure the longitudinal and Hall resistivity simultaneously and calculate the conductivity tensor ($\sigma_{ij}$).
 The zero-field critical temperature is defined by the maximum of $dR/dT$. The Presland formula \cite{PRESLAND199195} was used to derive the doping level, p, from the transition temperature. In  Fig.\ref{fig1}a we show the critical temperature  as a function of doping for all the samples used in this work. 
In Fig. \ref{fig1}b, we show the carrier concentration derived from the Hall resistivity measured at 120K as a function of p. The carrier density grows linearly with the doping level, p, that we extract from T$_c$. 
 
\begin{figure}
\includegraphics[width = 0.9\linewidth]{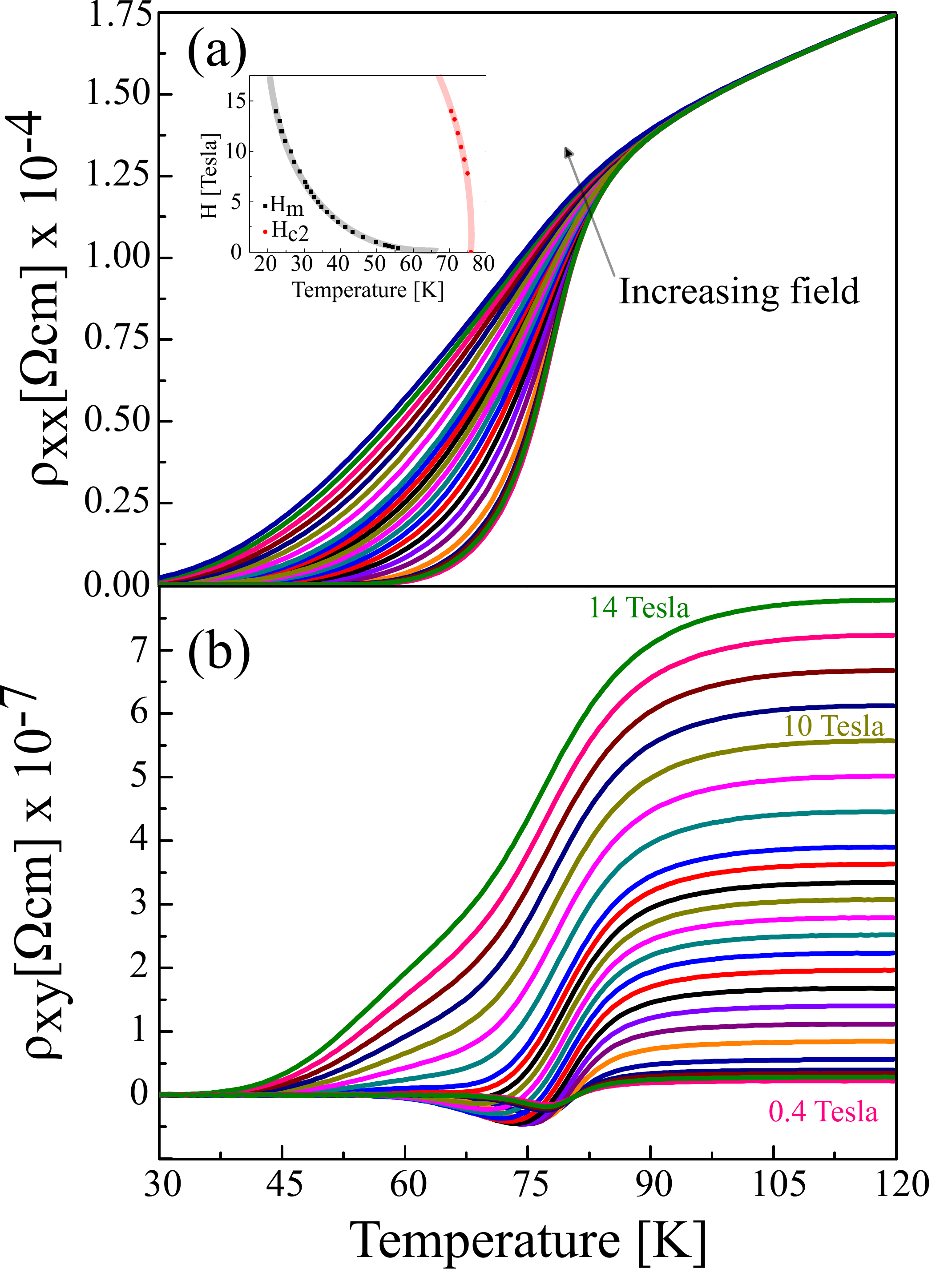}
\caption{ (a) Longitudinal resistivity and (b) Hall resistivity as a function of temperature for a T$_c$=77K (p=0.19) pristine Bi2212 sample for various magnetic fields between 0.4 and 14T. Inset of (a): Vortex lattice phase diagram for the same sample. Black dots show the vortex lattice melting line, red dots shows H$_{c2}$ (solid lines are guides to the eye). The flux flow regime lies between the lines.}
\label{fig2} 
\end{figure}

The magneto-resistance measurements were performed by applying a magnetic field at high temperature and measuring while cooling down. Then the temperature was raised again above T$_c$, the field polarity was reversed, and the measurement was repeated.
The field-cool procedure ensures that no vortices are trapped in the process of reversing the field. More details can be found in the supplementary materials\cite{supplementary}. 
We then separate the symmetric part (longitudinal resistivity) and the anti-symmetric part (Hall resistivity) with respect to the magnetic field. 
As an example, in Fig. \ref{fig2} we present the longitudinal and Hall resistivities for an overdoped  (p=0.19) T$_c$=77K pristine sample for magnetic fields ranging from 0.4T to 14T. The Hall resistivity at high temperatures varies slowly with temperature and is linear in magnetic field. Approaching the superconducting transition, the Hall resistivity drops. At low magnetic fields ( B$<$5T) the Hall resistivity changes sign for a range of temperatures around T$_c$. The Hall resistivity for all the samples can be found in the supplementary materials\cite{supplementary}. 

\section{Results}

\begin{figure}
\includegraphics[width = 1\linewidth]{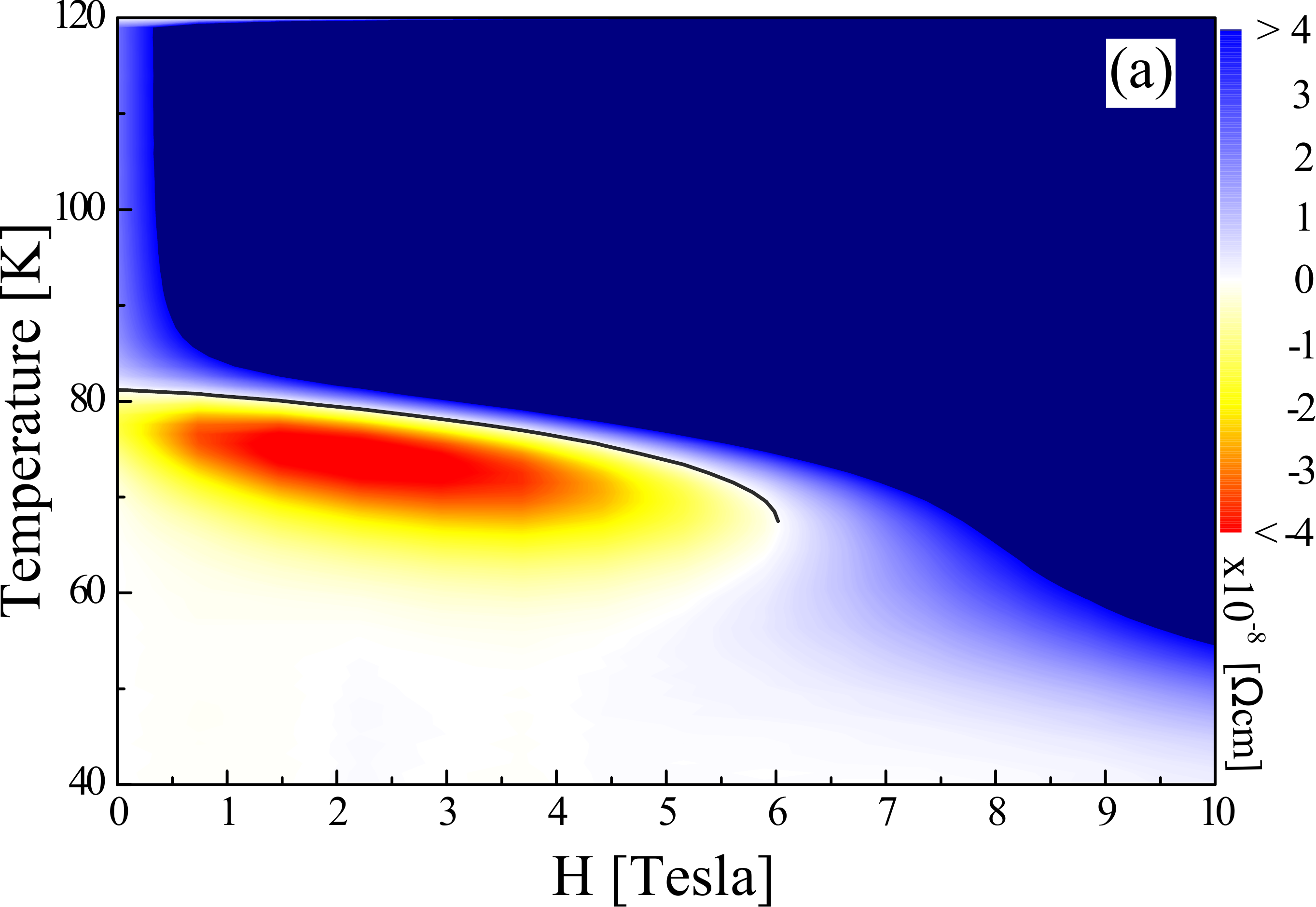}
\vfill
\includegraphics[width = 1\linewidth]{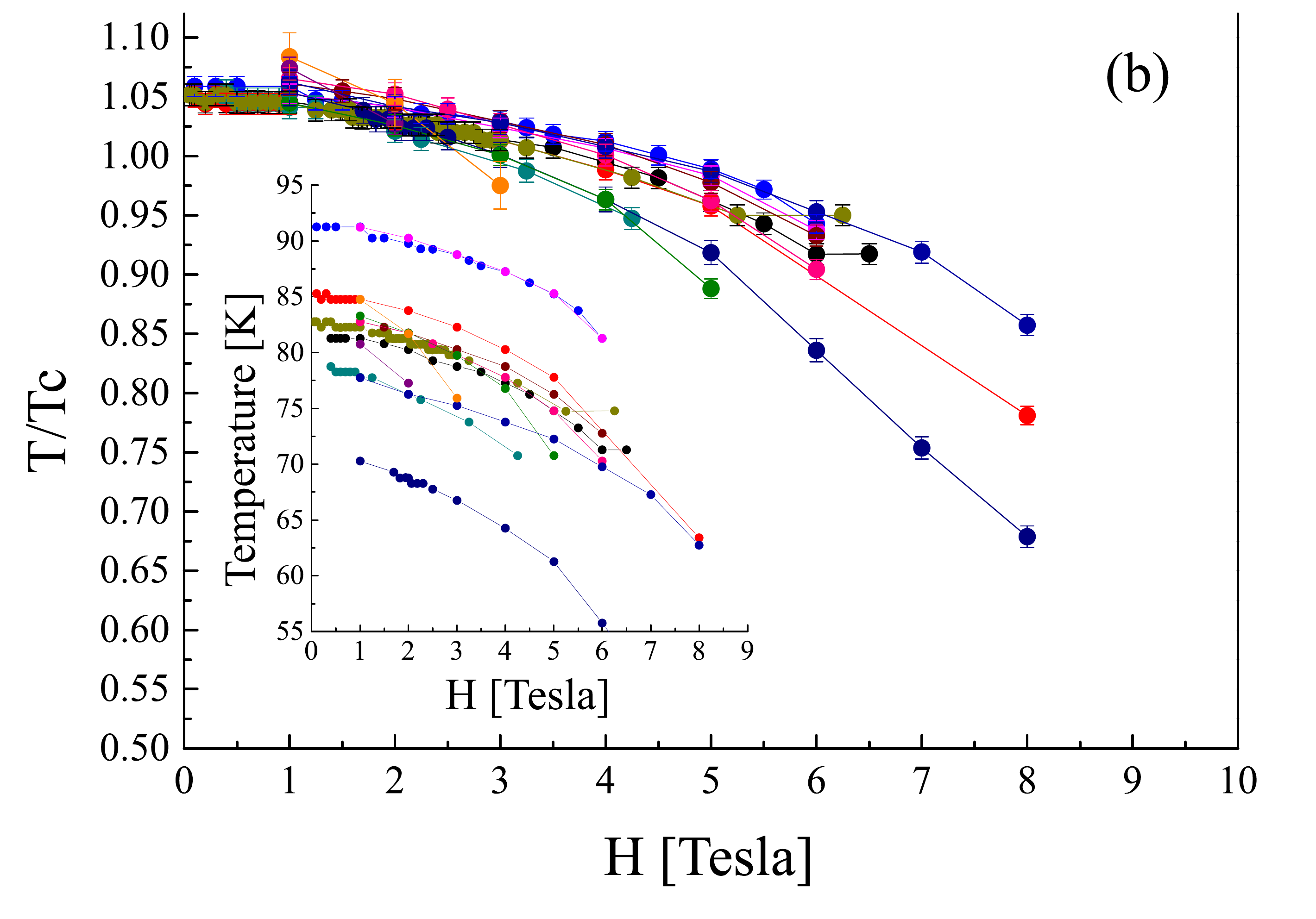}
\caption{(a) Hall resistivity as a function of temperature and magnetic field for a pristine overdoped T$_c$=77K sample. Black line represents the points at which the Hall resistivity changes sign. At low temperatures the vortices freeze and the Hall resistivity vanishes. (b) Hall sign reversal temperature normalized by the transition temperature as a function of the magnetic field for all the samples that exhibit a sign change. Inset: Zero Hall lines on an unnormlized temperature scale.}
\label{fig3}
\end{figure}

In  Fig. \ref{fig3}a we present the same Hall data shown in Fig. \ref{fig2}b as a color map. The anomaly is clearly seen as the red-yellow region in the plot. The solid black line tracks  $\rho_{xy}=0$, where the Hall changes sign. In \ref{fig3}b, we plot the sign reversal lines for all the samples that show a sign reversal, the temperature is normalized by the zero field T$_c$ for each sample. The inset shows the unnormalized data. At low magnetic fields all the normalized lines collapse one on top of the other. At higher magnetic fields there is some scattering and the sign reversal lines do not seem to scale together. The data indicates that in the limit of zero field the sign reversal takes place slightly above the transition temperature at about 1.05T$_c$.  

This scaling of the data over a large range of doping for samples with and without Zn substitution is remarkable, it shows that T$_c$ is the only temperature scale that plays a role in setting the Hall anomaly temperature. 
In any model describing the Hall conductance in the superconducting phase it is expected that the conductance of the vortex cores will play a role. A formation of any competing order that reconstructs the Fermi surface and creates electron pockets should be noticeable in the Hall data. Since it is not reasonable to assume that the transition temperature associated with the competing order coincides perfectly with T$_c$ over the entire phase diagram, we conclude that the competing phase, if exists, is not affecting the Hall signal. It is possible that in Bi2212 the CDW is very weak at low fields and does not lead to a significant Fermi surface reconstruction. Nonetheless, STM measurements done at similar fields to the ones used here find a clear "checkerboard" pattern surrounding the vortex cores \cite{HoffmanCores}. 

\begin{figure*}
\includegraphics[width=0.9\linewidth]{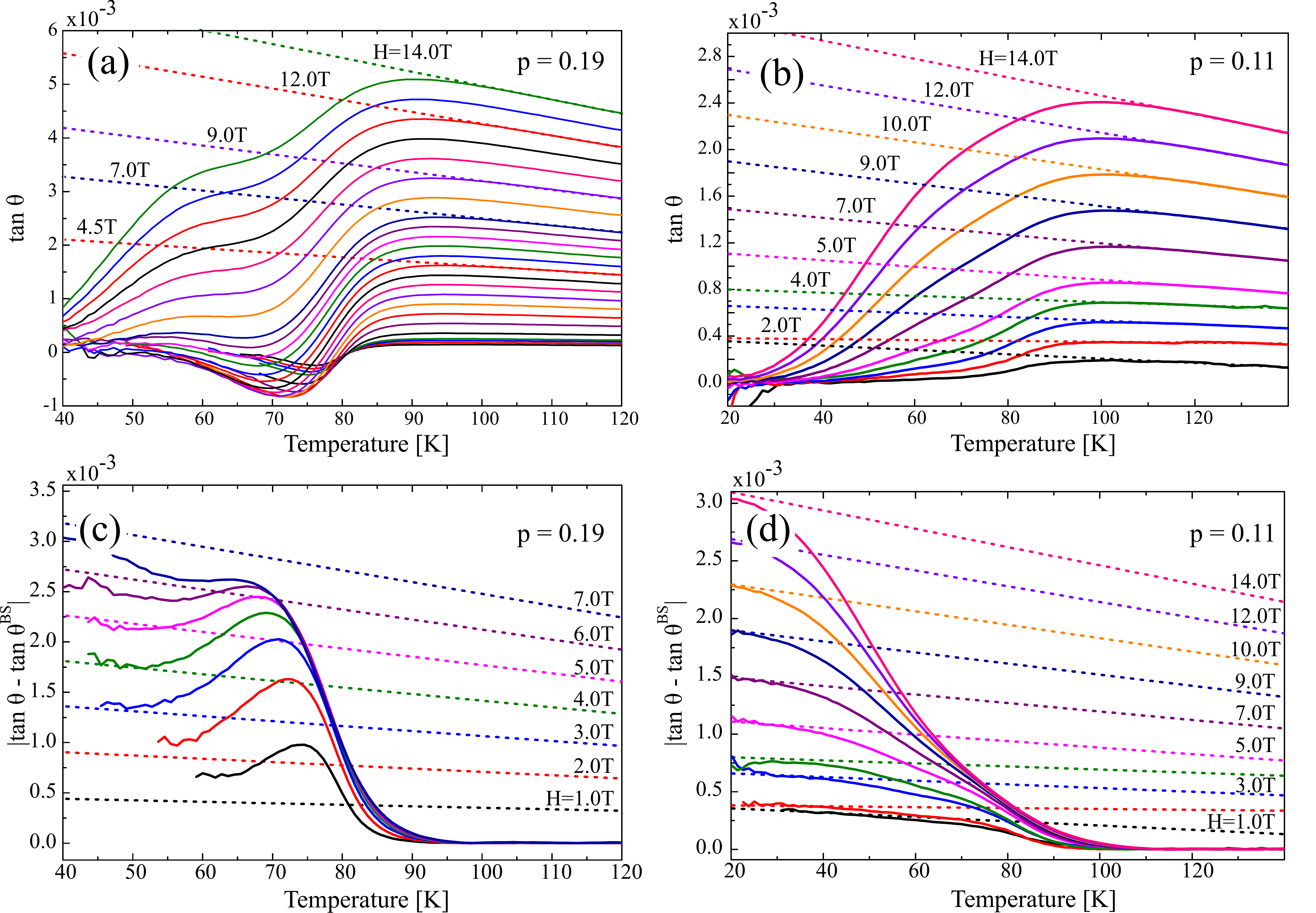}
\caption{Excess Hall voltage analysis for a Tc=77K p=0.19 sample and a Tc=71K p=0.11 sample. (a and b) Solid lines show the measured Hall angle as function of temperature for different magnetic fields. Dashed lines show the extrapolation of the high temperature data to lower temperatures. (c and d) The deviation of the data from the Bardeen-Stephen model predictions as a function of temperature for different magnetic fields (absolute value), compared to the Bardeen-Stephen model prediction.}

\label{bird}
\end{figure*}

Next, we provide a more quantitative description of the Hall conductivity in the flux flow regime.
To characterize the Hall anomaly we calculate the deviation of the Hall signal from the prediction of the Bardeen-Stephen model. 
According to this model:
$\rho_{i,j}^{Flux-flow}=\frac{H}{H_{c2}}\rho_{i,j}^{core}$\cite{bardeen1965theory}, where $i,j$ can be x or y, and $H_{c2}$ is the upper critical field at which superconductivity vanishes.
The result is that the Hall angle, $\tan\theta^{BS}=\frac{\rho_{yx}}{\rho_{xx}}=\tan\theta^{Normal}$, does not depend directly on the critical magnetic field, and should evolve continuously through the superconducting transition.  Using the Hall angle we bypassed the need to estimate the temperature dependence of $H_{c2}$. By extrapolating the normal state Hall angle to lower temperatures we can estimate the Bardeen-Stephen prediction for the Hall angle in the flux flow regime and compare it to our data. The flux flow regime is bounded between the vortex lattice melting line 
H$_m(T)$ \cite{Ramshaw2012} where vortices are free to move, and H$_{c2}(T)$ where vortices cease to exist (see Fig. \ref{fig2} Inset). We define H$_m$ to be the point at which the resistivity is 1000 times lower than the normal state resistivity.

We demonstrate the procedure used to calculate the deviation from the Bardeen-Stephen model in Fig. \ref{bird}, using data from the same overdoped T$_c$=77K sample, and an underdoped (p=0.11) T$_c$=71K sample which does not show a sign reversal. We first calculate the Hall angle vs temperature for different magnetic fields. In the normal state the Hall angle is linear in temperature so the extrapolation into the SC state is straightforward (shown as dashed lines).
 The difference between the measured Hall angle and the extrapolated values,  $\tan\theta^{excess} = \tan\theta-\tan\theta^{BS}$, are shown in Fig \ref{bird} c-d, along with the Bardeen-Stephen extrapolations.
  
 We emphasis that these results are only valid in a small range of temperature near the critical temperature. Deeper into the SC state the Hall angle extrapolation and the Bardeen-Stephen model  are not expected to describe the data correctly. 
 
 \section{discussion} 
 
The use of the conductivity data (instead of the resistivity data) allows us to treat the Hall voltage as a sum of different contributions, $\sigma_{xy} = \sigma_{xy}^{BS} + \sigma_{xy}^{excess}$,
where the excess Hall conductivity is given by: $\sigma_{xy}^{excess} = \tan \theta^{excess} \times\sigma_{xx}$.

We find that in addition to the Bardeen-Stephen term there is a contribution that is negative for any temperature and magnetic field.
Comparing the data of the two samples shown in Fig. \ref{bird}, we see that for the p=0.19 sample there is a range of temperatures and magnetic fields for which the negative excess Hall term is larger than the Bardeen-Stephen term resulting in a sign reversal. On the other hand, for the p=0.11 sample the negative term is smaller than the Bardeen-Stephen term for any temperature or magnetic field.

One possible source for the excess Hall conductivity are mobile charges trapped in the flux vortices. 
 Different models for vortex charge suggests different mechanisms for an excess Hall conductivity, which will result in different values and even different sign for a vortex charge\cite{khomskii1995charged,FEIGELMAN1994VortexCharge}.
 We adopt the model from Ref  \cite{Auerbach2019} : $$\sigma^{excess}_{xy}= \frac{2|e|Q_v}{h\cdot d}$$ where d is the c-axis lattice constant and $Q_v$ is a temperature and field dependent vortex charge per layer given in this model by:

 $$Q_v=Q_0 (\ln \frac{1}{H} + C(T))$$
  where $Q_0$ is the scale of the vortex charge which is expected to depend very weakly on the temperature around T$_c$ and C is a parameter that does not dependent on the magnetic field. 
 
\begin{figure}
\includegraphics[width= 1\linewidth]{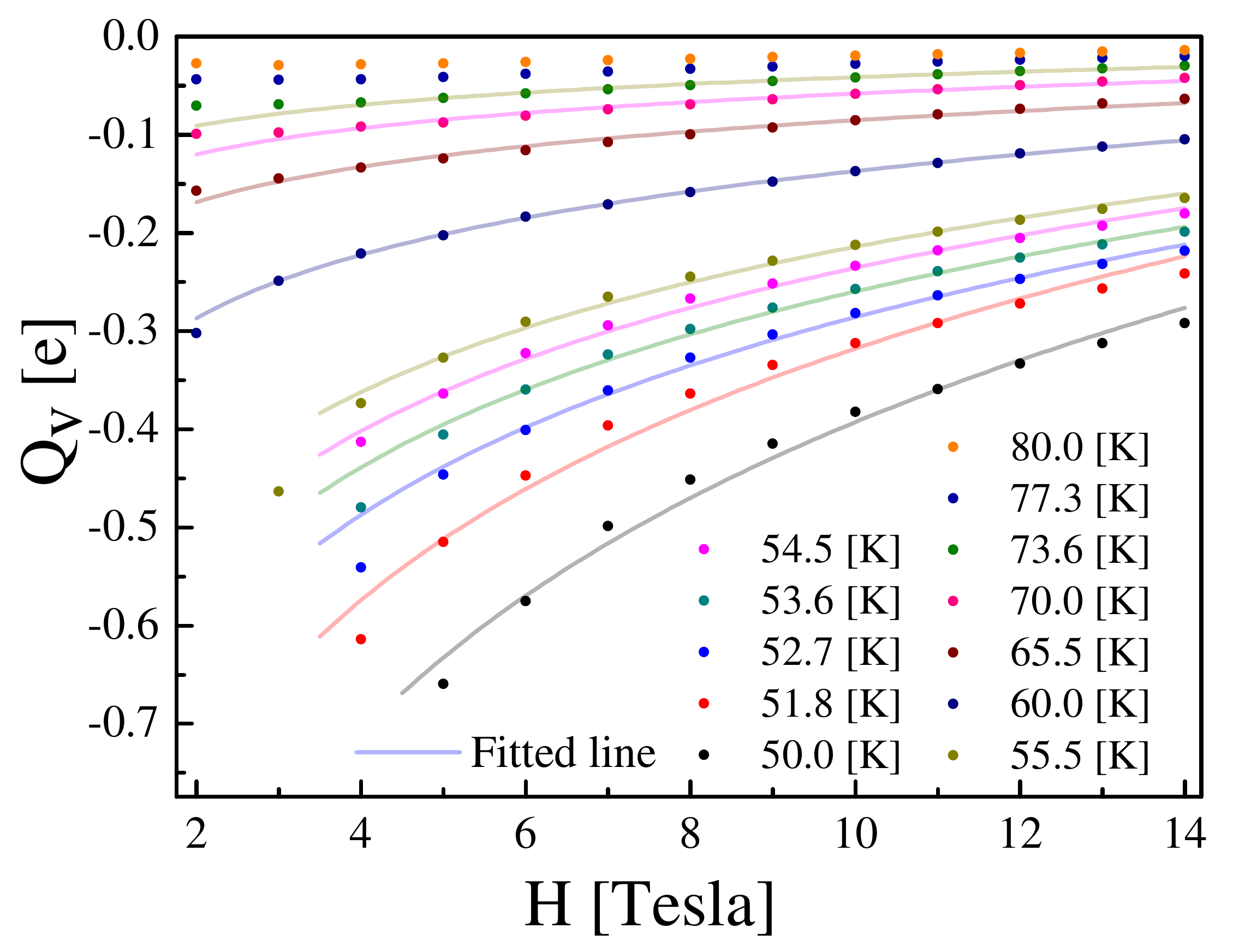}
\caption{\label{fit} Vortex charge per layer for an optimally doped 3\% Zn substituted sample as a function of magnetic field for various temperatures. Solid lines are best fits to the data using the vortex charge model.}
\end{figure}

 We used this expression to fit the excess Hall conductivity data. 
The fit is valid only in the vortex-liquid regime, where the vortices are mobile. This is a narrow temperature range around T$_c$ for low magnetic fields and an increasingly larger range of temperature as the magnetic field is increased. Within this range the model fits well the data as can be seen in Fig. \ref{fit} for a 3\% Zn optimally doped sample.  
Beyond this range, the freezing of the vortices will lead to a larger negative value of excess Hall conductivity, as we find for low temperatures and low magnetic fields.

We  apply our fitting procedure to all the measured samples.  
For each temperature we include only data points in the flux-flow regime. Overall the model fits well the data for all samples ( the fit results for all the samples are presented in the supplementary material\cite{supplementary}). 

The 
$\ln(\frac{1}{H})$ dependence of the data is not trivial. Other vortex-charge models with different field-dependence have been proposed\cite{FEIGELMAN1994VortexCharge}. Other models require a positive vortex charge in order for a Hall sign change to occur - we on the contrary, find a negative charge for the samples that show a sign reversal.

\begin{figure}[h]
    \centering
    \includegraphics[width=1\linewidth]{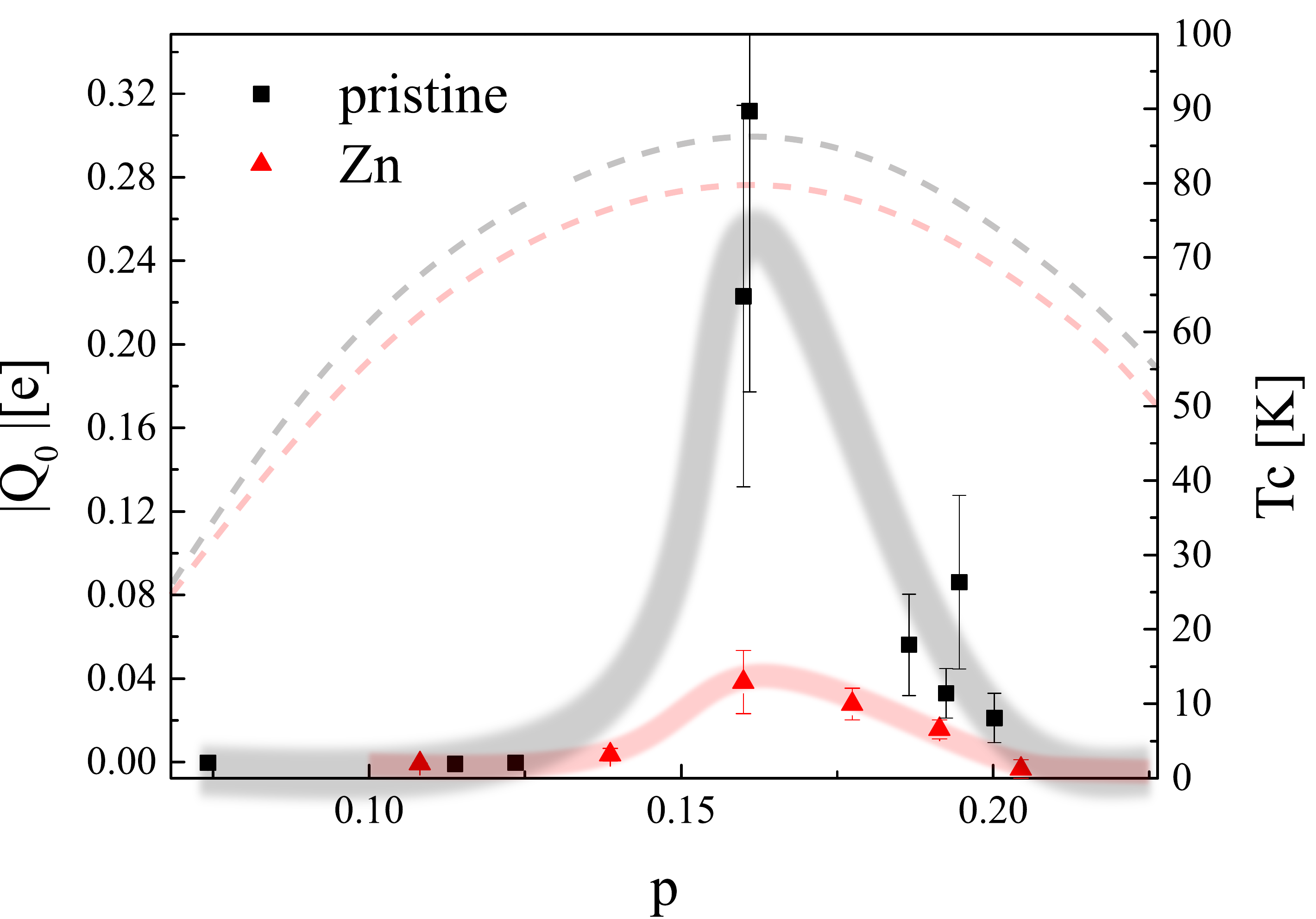}
    \caption{$|Q_0|$ as a function of the doping. The value presented is an average in the range of 0.8T$_c$ to T$_c$. Black squares are pristine samples, red triangles are 3\% Zn doped samples.  The solid lines are guides to the eye. The dashed lines represnt T$_c$.
    }
    \label{fig:chrage_mean}
\end{figure}

In Fig. \ref{fig:chrage_mean} we show $Q_0$ averaged in the range of (0.8-1.0)T$_c$ as a function of the doping. We look at a small range below the critical temperature where the extrapolation from the normal states is well justified. 

	 We find that the vortex charge strongly depends on the doping. It is  maximal at optimal doping and  decreases by about two orders of magnitude in the underdoped side.  The decrease in the overdoped side is  smaller.
    For the optimally doped pristine sample the vortex charge that we find is in agreement with the value found in \cite{Auerbach2019} using data from \cite{Zhao2019}.
The vortex charge scale for the optimmally doped sample is two orders of magnitude larger then estimated from a BCS analysis \cite{khomskii1995charged}. This is to be expected since the model based on BCS takes into account only charges which are located inside the vortex cores, and it uses a different screening model. In addition, this analysis predicts the vortex charge to be proportional to $\Delta^2/E_F$, where $\Delta$ is the superconducting gap and $E_F$ is the Fermi energy. For Bi2212, this should result in a vortex charge that is large in the underdoped side and decreases as the doping level increases, in contrary to our results.
 In addition, the vortex charge is significantly smaller in the Zn substituted samples compared to pristine samples while the gap does not change with Zn substitution \cite{Lubashevsky2011}.

\section{Summary}
We have measured the Hall and longitudinal resistivities as a function of magnetic field and temperature of Bi-2212 thin films at a range of doping levels and critical temperatures. We have observed a Hall sign reversal at doping levels down to $p \sim 0.12$. The sign reversal occurs at low magnetic fields ($B~<5T$) and always at temperatures around the superconducting critical temperature.  This suggests that the Hall sign reversal is closely related to superconductivity and it is not a result of a competing order. 

We then use the Bardeen-Stephen model as a starting point to calculate the excess Hall conductivity: $\sigma_{xy}=\sigma_{xy}^{BS}+\sigma_{xy}^{excess}$. We find that the excess Hall conductivity is negative for most temperatures and doping levels, not only where a sign reversal is observed. We find that a  vortex charge model \cite{Auerbach2019} fits very well our data. 
The model provides an interpretation to the excess Hall conductivity as a result of mobile charged vortices. We find that the vortex charge depends strongly on the doping level of the sample. The exact doping dependence of the vortex charge and the reason for the pronounced asymmetry in respect to optimal doping remain open questions.

\section{Acknowledgments} 
We acknowledge useful discussions with Assa Auerbach,  Vladimir Kalnitsky and Amit Keren.
This work was funded by the Israel Science Foundation under grant number 1263/21.
This research project was partially supported by the Helen Diller Quantum Center at the Technion.

\normalem
\bibliography{Hall_anomaly}

\end{document}